\documentclass[11pt,twoside]{article}
\usepackage{epsfig}

\begin{document}
\title{\bf Entanglement and Teleportation in Bipartite System}
\author{Satyabrata Adhikari \thanks{
satyabrata@bose.res.in}\\
\textsl{S. N. Bose National centre for Basic Sciences, Salt lake, Kolkata 700098, India}\\
 }
\date{}
\maketitle{} PACS numbers: 03.67.-a
\begin{abstract}
We present a mathematical formulation of old teleportation
protocol (original teleportation protocol introduced by Bennett
et.al.) for mixed state and study in detail the role of mixedness
of the two qubit quantum channel in a teleportation protocol. We
show that maximally entangled mixed state described by the density
matrix of rank-4 will be useful as a two qubit teleportation
channel to teleport a single qubit mixed state when the
teleportation channel parameter $p_{1}>\frac{1}{2}$. Also we
discuss the case when $p_{1}\leq \frac{1}{2}$.
\end{abstract}

\section{Introduction}
Bipartite system is nothing but a composite system consisting of
only two subsystems A and B situated distant apart. The relation
between the state of a composite quantum system as a whole and the
configuration of its parts is a very peculiar feature of quantum
information theory. It is not obvious that the reduced density
operator for system A is in any sense a description for the state
of system A. The physical justification for making this
identification is that the reduced density operator provides the
correct measurement statistics for measurements made on system A
\cite{nielsen}. Let us consider the two subsystems A and B
described by the Hilbert spaces $H_{A}$ and $H_{B}$ respectively,
then the composite system is described by the tensor product of
the Hilbert spaces $H_{A}$ and $H_{B}$ i.e., $H_{A} \otimes
H_{B}$. If the dimension of the Hilbert spaces $H_{A}$ and $H_{B}$
are $\textit{m}$ and $\textit{d}$ respectively then the dimension
of the composite system is $\textit{md}$. Equivalently, if the two
subsystems A and B described by the density matrices $\rho_{A}$
and $\rho_{B}$ are of order $\textit{m} \times \textit{m}$ and
$\textit{d}\times \textit{d} $ respectively then the composite
system is described by the density matrix $\rho_{AB}$ of order
$\textit{md}\times \textit{md}$. The density matrix $\rho_{AB}$
which describes the bipartite system can also be viewed as
$(\textit{md}\times \textit{md})$ ordered partitioned matrix which
consists of $\textit{m}^{2}$ (square of the dimension of the
Hilbert space $H_{A}$) block density matrices each of order $\textit{d}\times \textit{d}$.\\
The detection and characterization of entanglement even in the
bipartite system is a very difficult and challenging task in the
quantum information theory
\cite{aniello,agarwal,albeverio,dur,giovannetti,plesch}. Peres and
Horodecki family for the first time showed that inseparability of
bipartite composite system can be understood in terms of negative
eigenvalues of partial transpose of its density operator
\cite{horodecki,peres}. Peres-Horodecki criteria is necessary and
sufficient for $2\otimes2$ and $2\otimes3$ systems but it is only
sufficient for higher dimensional cases. Bipartite entangled state
can be used in various quantum information processing task like
quantum teleportation, superdense coding, quantum cryptography
etc. \cite{bennett1,bennett2,bennett3}\\
In this work our discussion will be restricted to only bipartite
system. Our main motivation of this work is to investigate whether
mixed teleportation channel will be useful to teleport a single qubit mixed state. \\
The paper is organised as follows. In section-2, we study the
properties of semi-definite matrices. In section-3, we study the
conventional teleportation scheme for mixed state and show that
even mixed state as a teleportation channel will be useful to
teleport a single qubit mixed state. In section-4, conclusions are
drawn.

\section{A Few properties of density matrices in terms of block matrices}
Let $\rho_{ij} \in F^{d\times d}$ $\forall$ i,j=0,1,2,...,d-1
denote the block matrices. We define the bipartite composite
system in terms of block matrices as
\begin{eqnarray} [\rho_{AB}] =
\left(\begin{matrix}{[\rho_{00}] & [\rho_{01}] &  [\rho_{02}] &
.... & [\rho_{0d-1}] \cr [\rho_{01}]^{\dagger} & [\rho_{11}] &
[\rho_{12}] & .... & [\rho_{1d-1}] \cr [\rho_{02}]^{\dagger} &
[\rho_{12}]^{\dagger} & [\rho_{22}] & .... & [\rho_{2d-1}] \cr
.... & .... & .... & .... & ....  \cr [\rho_{0d-1}]^{\dagger} &
[\rho_{1d-1}]^{\dagger} & [\rho_{2d-1}]^{\dagger} & .... &
[\rho_{d-1d-1}] }\end{matrix}\right) \label{blkmat}
\end{eqnarray}
For instance, if we consider $2\otimes 3$ system then the whole
system is described by a $6\times6$ ordered partition matrix which
consist of $3\times 3$ ordered four block matrices. In the matrix
form, it is given by
\begin{eqnarray} [\rho_{AB}] =
\left(\begin{matrix}{[\rho_{00}] & [\rho_{01}] \cr
[\rho_{01}^{\dagger}] & [\rho_{11}] }\end{matrix}\right)
\label{blkmatex}
\end{eqnarray}
where $[\rho_{00}]=\left(\begin{matrix}{a_{1} & a_{2} &  a_{3} &
 \cr a_{2}^{*} & a_{4} & a_{5} \cr a_{3}^{*} & a_{5}^{*} &
 a_{6}}\end{matrix}\right)$, $[\rho_{01}]= \left(\begin{matrix}{ c_{1} & c_{2} &
c_{3} \cr c_{4} & c_{5} & c_{6} \cr  c_{7} & c_{8} & c_{9}
}\end{matrix}\right)$ and $[\rho_{11}]=\left(\begin{matrix}{ b_{1}
& b_{2} & b_{3} \cr  b_{2}^{*} & b_{4} & b_{5} \cr b_{3}^{*} &
b_{5}^{*} & b_{6}}\end{matrix}\right)$.\\\\
\textbf{P1.} The density matrix $[\rho_{AB}]$ is positive
semi-definite iff the diagonal block density matrices $[\rho_{ii}]$ (i=0,1,....d-1) are positive semi-definite.\\
\textbf{P2.} If the density matrix $[\rho_{AB}]$ is
positive-semidefinite then
\begin{eqnarray} det
\left(\begin{matrix}{det[\rho_{00}] & det[\rho_{01}] &
det[\rho_{02}] & .... & det[\rho_{0d-1}] \cr
det[\rho_{01}]^{\dagger} & det[\rho_{11}] & det[\rho_{12}] & ....
& det[\rho_{1d-1}] \cr det[\rho_{02}]^{\dagger} &
det[\rho_{12}]^{\dagger} & det[\rho_{22}] & .... &
det[\rho_{2d-1}] \cr .... & .... & .... & .... & ....  \cr
det[\rho_{0d-1}]^{\dagger} & det[\rho_{1d-1}]^{\dagger} &
det[\rho_{2d-1}]^{\dagger} & .... & det[\rho_{d-1d-1}]
}\end{matrix}\right)\leq det[\rho_{AB}] \label{P2(i)}
\end{eqnarray}
and
\begin{eqnarray}
\left(\begin{matrix}{tr[\rho_{00}] & tr[\rho_{01}] & tr[\rho_{02}]
& .... & tr[\rho_{0d-1}] \cr tr[\rho_{01}]^{\dagger} &
tr[\rho_{11}] & tr[\rho_{12}] & .... & tr[\rho_{1d-1}] \cr
tr[\rho_{02}]^{\dagger} & tr[\rho_{12}]^{\dagger} & tr[\rho_{22}]
& .... & tr[\rho_{2d-1}] \cr .... & .... & .... & .... & ....  \cr
tr[\rho_{0d-1}]^{\dagger} & tr[\rho_{1d-1}]^{\dagger} &
tr[\rho_{2d-1}]^{\dagger} & .... & tr[\rho_{d-1d-1}]
}\end{matrix}\right) \textit{is positive semi-definite}
\label{P2(ii)}
\end{eqnarray}
\textbf{P3.} If the density matrix $[\rho_{AB}]$ is
positive-semidefinite and represents $2\otimes d$ system i.e. if
$[\rho_{AB}]$ is of the form
\begin{eqnarray} [\rho_{AB}] =
\left(\begin{matrix}{[\rho_{00}] & [\rho_{01}] \cr
[\rho_{01}^{\dagger}] & [\rho_{11}] }\end{matrix}\right)
\label{blkmat1}
\end{eqnarray}  then
\begin{eqnarray}
tr([\rho_{01}^{\dagger}][\rho_{01}]) \leq
\sqrt{tr[\rho_{00}^{2}]tr[\rho_{11}^{2}]}\leq
tr[\rho_{00}]tr[\rho_{11}] \label{P3(i)}
\end{eqnarray}
and
\begin{eqnarray}
0\leq det[\rho_{00}]det[\rho_{11}]-|det[\rho_{01}]|^{2}\leq
det[\rho_{AB}]\leq det[\rho_{00}]det[\rho_{11}]\label{P3(ii)}
\end{eqnarray}
\section{Bennett's et.al. teleportation scheme for mixed states}
Quantum teleportation \cite{bennett1} is a fundamental and vital
quantum information processing task where an unknown quantum state
is transferred from a sender (Alice) to a distant receiver (Bob)
using their prior shared entangled state and two bits of classical
communication. In the standard teleportation scheme for moving
unknown state of a qubit, Alice first performs Bell state
measurement on the combined system of input qubit and her half of
a maximally entangled pure state. She then uses classical
communication channel to tell the measurement result to Bob. After
getting the measurement result, Bob applies unitary transformation
on his qubit to complete the teleportation protocol.\\
In this section we provide the mathematical formulation (in the
light of block matrices) of the conventional teleportation scheme
of a qubit by taking the more general two qubit teleportation
channel. This formulation will help us in studying the original
teleportation protocol in a easier way even with single qubit
mixed input state and two qubit mixed teleportation channel. We
now proceed step by step to develop our mathematical formulation for teleportation protocol.\\\\
\textbf{Step-I:} Let us consider a qubit to be teleported
described by the density matrix
\begin{eqnarray}
[\rho_{1}]= \left(\begin{matrix}{\rho_{0,0} & \rho_{0,1} \cr
\rho_{0,1}^{*} & \rho_{1,1} }\end{matrix}\right) \label{rho1}
\end{eqnarray}
To proceed with the teleportation scheme, we need a quantum
channel through which Alice sends her message encoding the state
(\ref{rho1}) to Bob. Thus the two-qubit quantum channel needed to
teleport a single qubit (\ref{rho1}) is described by the density
matrix
\begin{eqnarray}
[\rho_{23}]= \left(\begin{matrix}{\rho_{00,00} & \rho_{00,01} &
\rho_{00,10} & \rho_{00,11} \cr \rho_{00,01}^{*} & \rho_{01,01} &
\rho_{01,10} & \rho_{01,11} \cr \rho_{00,10}^{*} &
\rho_{01,10}^{*} &  \rho_{10,10} & \rho_{10,11} \cr
\rho_{00,11}^{*} & \rho_{01,11}^{*} & \rho_{10,11}^{*} &
\rho_{11,11} }\end{matrix}\right) \label{rho23}
\end{eqnarray}
The qubits 1 and 2 are possessed by Alice and qubit 3 by Bob.\\
\textbf{Step-II:} Next we combine the single qubit system
described by the density operator $[\rho_{1}]$ and the two-qubit
system described by the density operator $[\rho_{23}]$ using
tensor product between the two systems. As a result the composite
three qubit system is given by
\begin{eqnarray}
[\rho_{1}]\otimes[\rho_{23}]= \left(\begin{matrix}{\rho_{000,000}
& \rho_{000,001} & \rho_{000,010} & \rho_{000,011} &
\rho_{000,100} & \rho_{000,101} & \rho_{000,110} & \rho_{000,111}
\cr \rho_{000,001}^{*} & \rho_{001,001} & \rho_{001,010} &
\rho_{001,011} & \rho_{001,100} & \rho_{001,101} & \rho_{001,110}
& \rho_{001,111} \cr \rho_{000,010}^{*} & \rho_{001,010}^{*} &
\rho_{010,010} & \rho_{010,011} & \rho_{010,100} & \rho_{010,101}
& \rho_{010,110} & \rho_{010,111} \cr \rho_{000,011}^{*} &
\rho_{001,011}^{*} & \rho_{010,011}^{*} & \rho_{011,011} &
\rho_{011,100} & \rho_{011,101} & \rho_{011,110} & \rho_{011,111}
\cr \rho_{000,100}^{*} & \rho_{001,100}^{*} & \rho_{010,100}^{*} &
\rho_{011,100}^{*} & \rho_{100,100} & \rho_{100,101} &
\rho_{100,110} & \rho_{100,111} \cr \rho_{000,101}^{*} &
\rho_{001,101}^{*} & \rho_{010,101}^{*} & \rho_{011,101}^{*} &
\rho_{100,101}^{*} & \rho_{101,101} & \rho_{101,110} &
\rho_{101,111} \cr \rho_{000,110}^{*} & \rho_{001,110}^{*} &
\rho_{010,110}^{*} & \rho_{011,110}^{*} & \rho_{100,110}^{*} &
\rho_{101,110}^{*} & \rho_{110,110} & \rho_{110,111} \cr
\rho_{000,111}^{*} & \rho_{001,111}^{*} & \rho_{010,111}^{*} &
\rho_{011,111}^{*} & \rho_{100,111}^{*} & \rho_{101,111}^{*} &
\rho_{110,111}^{*} & \rho_{111,111} }\end{matrix}\right)
\label{rho123}
\end{eqnarray}
In the forthcoming step, Alice has to perform Bell-state
measurement on her qubits 1 and 2. To do the Bell-state
measurement, Alice transforms the basis of the qubits in her side
from the computational basis $\{|00>, |01>, |10>, |11>\}$ to the
Bell basis $\{|\Phi^{+}>, |\Phi^{-}>, |\Psi^{+}>, |\Psi^{-}>\}$,
where $|\Phi^{\pm}>=\frac{1}{\sqrt{2}}(|00>\pm|11>)$ and
$|\Psi^{\pm}>=\frac{1}{\sqrt{2}}(|01>\pm|10>)$. Therefore, when
Alice's qubits are written in the Bell basis and Bob's qubit in
the computational basis, the eqn. (\ref{rho123}) looks like
\begin{eqnarray}
[\rho_{1}]\otimes[\rho_{23}]= |\Phi^{+}>_{12}<\Phi^{+}|\otimes
\left(\begin{matrix}{\rho_{000,000}+\rho_{000,110} &&
\rho_{000,001}+\rho_{000,111} {}\nonumber\\
+\rho_{110,000}+\rho_{110,110} && +\rho_{110,001}+\rho_{110,111}
\cr \rho_{000,001}^{*}+ \rho_{000,111}^{*} &&
\rho_{001,001}+\rho_{001,111}{}\nonumber\\+\rho_{110,001}^{*}+
\rho_{110,111}^{*} && +\rho_{111,001}+ \rho_{111,111}
}\end{matrix}\right)
{}\nonumber\\
+|\Phi^{-}>_{12}<\Phi^{-}|\otimes
\left(\begin{matrix}{\rho_{000,000}-\rho_{000,110} &&
\rho_{000,001}-\rho_{000,111} {}\nonumber\\
-\rho_{110,000}+\rho_{110,110} && -\rho_{110,001}+\rho_{110,111}
\cr \rho_{000,001}^{*}- \rho_{000,111}^{*} &&
\rho_{001,001}-\rho_{001,111}{}\nonumber\\-\rho_{110,001}^{*}+
\rho_{110,111}^{*} && -\rho_{111,001}+ \rho_{111,111}
}\end{matrix}\right)
{}\nonumber\\
+|\Psi^{+}>_{12}<\Psi^{+}|\otimes
\left(\begin{matrix}{\rho_{010,010}+\rho_{010,100} &&
\rho_{010,011}+\rho_{010,101} {}\nonumber\\
+\rho_{100,010}+\rho_{100,100} && +\rho_{100,011}+\rho_{100,101}
\cr \rho_{010,011}^{*}+\rho_{010,101}^{*} &&
\rho_{011,011}+\rho_{011,101}{}\nonumber\\\rho_{100,011}^{*}+\rho_{100,101}^{*}
&& \rho_{101,011}+ \rho_{111,111} }\end{matrix}\right)
{}\nonumber\\
+|\Psi^{-}>_{12}<\Psi^{-}|\otimes
\left(\begin{matrix}{\rho_{010,010}+\rho_{010,100} &&
\rho_{010,011}-\rho_{010,101} {}\nonumber\\
+\rho_{100,010}+\rho_{100,100} && -\rho_{100,011}+\rho_{100,101}
\cr \rho_{010,011}^{*}-\rho_{010,101}^{*} &&
\rho_{011,011}-\rho_{011,101}{}\nonumber\\-\rho_{100,011}^{*}+\rho_{100,101}^{*}
&& -\rho_{101,011}+ \rho_{111,111} }\end{matrix}\right)
{}\nonumber\\
+~ \textrm{other terms containing}~~|\Psi^{+}>_{12}<\Psi^{-}|,~~
|\Psi^{-}>_{12}<\Psi^{+}|~~ \textrm{etc.}~~~~~~~~~~~~
\label{rho123n}
\end{eqnarray}
\textbf{Step-III:} Alice performs Bell state measurement on her
qubit and sends the measurement result to Bob by expending two
classical bits.\\\\
\textbf{Step-IV:} After getting the measurement result, Bob
operates with some unitary operator on the received state to
retrieve the state sent by Alice.\\\\
\textbf{Illustration}\\
Let us understand the above steps by taking a specific
example:\\
Suppose an arbitrary input state that Alice wants to teleport is
described by the density matrix
\begin{eqnarray} [\rho_{1}]= \left(\begin{matrix}{x & y \cr y^{*}
& 1-x }\end{matrix}\right) \label{rho1ex}
\end{eqnarray}
Here the input parameters $x$ and $y$ are unknown to Alice.\\
Let the two-qubit teleportation channel shared by two distant
partners Alice and Bob be given by
\begin{eqnarray}
[\rho_{23}]= \left(\begin{matrix}{a & 0 & 0 & e \cr 0 & b & c & 0
\cr 0 & c^{*} & d & 0 \cr e^{*} & 0 & 0 & 1-a-b-d
}\end{matrix}\right) \label{rho23ex}
\end{eqnarray}
The composite three qubit system $[\rho_{1}]\otimes[\rho_{23}]$
can be written as
\begin{eqnarray}
\left(\begin{matrix}{ax & 0 & 0 & xe & ya & 0 & 0 & ye \cr 0 & xb
& xc & 0 & 0 & yb & yc & 0 \cr 0 & c^{*}x & dx & 0 & 0 & yc^{*} &
yd & 0 \cr xe^{*} & 0 & 0 & x(1-a-b-d) & ye^{*} & 0 & 0 &
y(1-a-b-d) \cr ay^{*} & 0 & 0 & ey^{*} & (1-x)a & 0 & 0 & e(1-x)
\cr 0 & by^{*} & cy^{*} & 0 & 0 & (1-x)b & (1-x)c & 0 \cr 0 &
y^{*}c^{*} & dy^{*} & 0 & 0 & (1-x)c^{*} & (1-x)d & 0 \cr
y^{*}e^{*} & 0 & 0 & (1-a-b-d)y^{*} & (1-x)e^{*} & 0 & 0 &
(1-x)(1-a-b-d) }\end{matrix}\right) \label{rho123ex}
\end{eqnarray}
Using eqn.(\ref{rho123n}), eqn. (\ref{rho123ex}) can be rewritten
as
\begin{eqnarray}
[\rho_{1}]\otimes[\rho_{23}]= |\Phi^{+}>_{12}<\Phi^{+}|\otimes
\left(\begin{matrix}{xa+(1-x)d & y^{*}c^{*}+ye \cr yc+y^{*}e^{*} &
xb+(1-x)(1-a-b-d) }\end{matrix}\right)
{}\nonumber\\
+|\Phi^{-}>_{12}<\Phi^{-}|\otimes \left(\begin{matrix}{xa+(1-x)d &
-y^{*}c^{*}-ye \cr -yc-y^{*}e^{*} & xb+(1-x)(1-a-b-d)
}\end{matrix}\right)
{}\nonumber\\
+|\Psi^{+}>_{12}<\Psi^{+}|\otimes \left(\begin{matrix}{ xd+(1-x)a&
 yc^{*}+y^{*}e \cr cy^{*}+ye^{*} & x(1-a-b-d)+(1-x)b }\end{matrix}\right)
{}\nonumber\\
+|\Psi^{-}>_{12}<\Psi^{-}|\otimes \left(\begin{matrix}{ xd+(1-x)a&
 -yc^{*}-y^{*}e \cr -cy^{*}-ye^{*} & x(1-a-b-d)+(1-x)b
}\end{matrix}\right)
{}\nonumber\\
+~ \textrm{other terms containing}~~|\Psi^{+}>_{12}<\Psi^{-}|,~~
|\Psi^{-}>_{12}<\Psi^{+}|~~ \textrm{etc.}~~~ \label{rho123m}
\end{eqnarray}
Thereafter Alice performs Bell state measurement on her qubit and sends the measurement result to Bob.\\
If Alice's measurement outcome is $|\Phi^{+}>$ or $|\Phi^{-}>$ or
$|\Psi^{+}>$ or $|\Psi^{-}>$ then the corresponding states
received by Bob are described by the density matrices
\begin{eqnarray}
&&[\rho^{|\Phi^{\pm}>}_{B}]=(1/N)\left(\begin{matrix}{xa+(1-x)d &
\pm (y^{*}c^{*}+ye) \cr \pm (yc+y^{*}e^{*}) & xb+(1-x)(1-a-b-d)
}\end{matrix}\right) {}\nonumber\\&&
\textrm{or}~~~[\rho^{|\Psi^{\pm}>}_{B}]=(1/N_{1})\left(\begin{matrix}{
xd+(1-x)a& \pm (yc^{*}+y^{*}e) \cr \pm (cy^{*}+ye^{*}) &
x(1-a-b-d)+(1-x)b }\end{matrix}\right)\label{B1}
\end{eqnarray}
where $N=x(a+b)+(1-x)(1-a-b)$ and $N_{1}=x(1-a-b)+(1-x)(a+b)$ are
the normalization constants.\\
Now before going to fourth step of the teleportation protocol, we
want to simplify the situation and consider the maximally
entangled mixed state (MEMS) \cite{ishizaka,wei} as a two qubit
teleportation channel shared by Alice and Bob. Therefore the two
qubit teleportation channel (\ref{rho23ex}) reduces to
\begin{eqnarray}
[\rho_{23}]^{MEMS}= \left(\begin{matrix}{\frac{p_{1}+p_{3}}{2} & 0
& 0 & \frac{p_{3}-p_{1}}{2} \cr 0 & p_{2} & 0 & 0 \cr 0 & 0 &
p_{4} & 0 \cr \frac{p_{3}-p_{1}}{2} & 0 & 0 &
1-p_{2}-p_{4}-\frac{p_{1}+p_{3}}{2}}\end{matrix}\right)
\label{rho23ex1}
\end{eqnarray}
where $p_{1}\geq p_{2} \geq p_{3} \geq p_{4}$ and $p_{1}+p_{2}+
p_{3}+p_{4}=1$.\\\\
\textbf{CASE-I}: When $rank([\rho_{23}]^{MEMS})=1$ i.e., when
$p_{2}=p_{3}=p_{4}=0$.\\
In this case $[\rho_{23}]^{MEMS}$ reduces to maximally entangled
pure state (MEPS). Therefore,
\begin{eqnarray}
[\rho_{23}]^{MEPS}= \left(\begin{matrix}{\frac{1}{2} & 0 & 0 &
\frac{-1}{2} \cr 0 & 0 & 0 & 0 \cr 0 & 0 & 0 & 0 \cr \frac{-1}{2}
& 0 & 0 & \frac{1}{2}}\end{matrix}\right) \label{rho23meps}
\end{eqnarray}
The maximum concurrence \cite{wootters} of the state described by
the density matrix $[\rho_{23}]^{MEPS}$ is unity.\\
If Alice uses $[\rho_{23}]^{MEPS}$ as a teleportation channel to
teleport an unknown qubit described by the density matrix
(\ref{rho1ex}) then the state after Alice's measurement is given
by
\begin{eqnarray}
[\rho^{|\Phi^{\pm}>}_{B_{1}}]= \left(\begin{matrix}{x & \mp y \cr
\mp y^{*} & 1-x}\end{matrix}\right) ~~~~\textrm{and}~~~~~
[\rho^{|\Psi^{\pm}>}_{B_{2}}]= \left(\begin{matrix}{1-x & \mp
y^{*} \cr \mp y & x}\end{matrix}\right) \label{B12}
\end{eqnarray}
According to the measurement results $|\Phi^{+}>$, $|\Phi^{-}>$,
$|\Psi^{+}>$ or $|\Psi^{-}>$ sent by Alice with the help of two
classical bits, Bob operates with unitary operators $\sigma_{z}$,
$I$, $\sigma_{y}$ and $\sigma_{x}$ on his received qubit to
retrieve the original state undistorted.\\\\
\textbf{CASE-II}: When $rank([\rho_{23}]^{MEMS})=2$ i.e., when
$p_{3}=p_{4}=0$. In this case $[\rho_{23}]^{MEMS}$ reduces to
\begin{eqnarray}
[\rho_{23}]^{MEMS}_{r2}= \left(\begin{matrix}{\frac{p_{1}}{2} & 0
& 0 & \frac{-p_{1}}{2} \cr 0 & 1-p_{1} & 0 & 0 \cr 0 & 0 & 0 & 0
\cr \frac{-p_{1}}{2} & 0 & 0 & \frac{p_{1}}{2}}\end{matrix}\right)
\label{rho23r2}
\end{eqnarray}
The maximum concurrence of the state $[\rho_{23}]^{MEMS}_{r2}$ is $C([\rho_{23}]^{MEMS}_{r2})=p_{1}$. \\
If Alice uses $[\rho_{23}]^{MEMS}_{r2}$ as a teleportation channel
to teleport an unknown qubit described by the density matrix
(\ref{rho1ex}) then the state received by Bob after Alice's
measurement is given by
\begin{eqnarray}
&&[\rho^{|\Phi^{\pm}>}_{B_{3}}]=
\frac{1}{N}\left(\begin{matrix}{\frac{xp_{1}}{2} & \mp
\frac{yp_{1}}{2} \cr \mp \frac{y^{*}p_{1}}{2} &
x(1-p_{1})+\frac{(1-x)p_{1}}{2}}\end{matrix}\right) ~\textrm{and}~
{}\nonumber\\&&[\rho^{|\Psi^{\pm}>}_{B_{4}}]=
\frac{1}{N_{1}}\left(\begin{matrix}{\frac{(1-x)p_{1}}{2} & \mp
\frac{y^{*}p_{1}}{2} \cr \mp \frac{yp_{1}}{2} &
\frac{xp_{1}}{2}+(1-x)(1-p_{1})}\end{matrix}\right) \label{B34}
\end{eqnarray}
where $N=x(1-\frac{p_{1}}{2})+(1-x)\frac{p_{1}}{2}$ and $N_{1}=x\frac{p_{1}}{2}+(1-x)(1-\frac{p_{1}}{2})$.\\
Based on measurement results $|\Phi^{+}>$, $|\Phi^{-}>$,
$|\Psi^{+}>$ or $|\Psi^{-}>$, Bob operates with unitary operators
$\sigma_{z}$, $I$, $\sigma_{y}$ and $\sigma_{x}$ respectively on
his received qubit and therefore the state given in (\ref{B34})
reduces to
\begin{eqnarray}
\sigma_{z}[\rho^{|\Phi^{+}>}_{B}]\sigma_{z}^{\dagger}=I[\rho^{|\Phi^{-}>}_{B}]I
\equiv [\rho^{B_{3}}_{3}]=
\frac{1}{N}\left(\begin{matrix}{\frac{xp_{1}}{2} &
\frac{yp_{1}}{2} \cr \frac{y^{*}p_{1}}{2} &
x(1-p_{1})+\frac{(1-x)p_{1}}{2}}\end{matrix}\right) ~\textrm{and}~
{}\nonumber\\\sigma_{y}[\rho^{|\Psi^{+}>}_{B}]\sigma_{y}^{\dagger}=\sigma_{x}[\rho^{|\Psi^{-}>}_{B}]\sigma_{x}^{\dagger}
\equiv [\rho^{B_{4}}_{3}]=
\frac{1}{N_{1}}\left(\begin{matrix}{\frac{xp_{1}}{2}+(1-x)(1-p_{1})
&  \frac{yp_{1}}{2} \cr \frac{y^{*}p_{1}}{2} &
\frac{(1-x)p_{1}}{2}}\end{matrix}\right) \label{rhoB34}
\end{eqnarray}
From (\ref{rhoB34}) it is clear that even after unitary
transformation Bob could not retrieve the sent state quite
properly. Thus to see how good the input state is teleported we
have to calculate the square of the Hilbert Schmidt norm
\cite{buzek} of the difference between two density matrices
$[\rho_{1}]$ at Alice's side and $[\rho^{B_{3}}_{3}]~~
\textrm{or}~~
[\rho^{B_{4}}_{3}]$ (corresponding to different measurements) at Bob's side.\\
The Hilbert Schmidt norm is given by
\begin{eqnarray}
&&D_{1}= Tr([\rho_{1}]-[\rho_{3}^{B_{3}}])^{2}=
\frac{2}{N^{2}}(x^{2}(1-p_{1})^{2}(x^{2}+|y|^{2}))~~~~\textrm{and}~~~~
{}\nonumber\\&& D_{2}= Tr([\rho_{1}]-[\rho_{3}^{B_{4}}])^{2}=
\frac{2}{N^{2}_{1}}((1-x)^{2}(1-p_{1})^{2}((1-x)^{2}+|y|^{2}))
\label{dist1}
\end{eqnarray}
Here we note that if we put $p_{1}=1$ in (\ref{dist1}) then we
obtain the result as in case-I (i.e., rank-1 case) where the state
(\ref{rho1ex}) is perfectly teleported (i.e., without introducing
any error) via quantum channel (\ref{rho23meps}).\\
Let us consider a class of input states given by
\begin{eqnarray} [\rho_{1}^{in}]= \left(\begin{matrix}{\frac{1}{2} & y \cr y^{*}
& \frac{1}{2} }\end{matrix}\right) \label{rho1inex}
\end{eqnarray}
Here also the input parameter $y$ is completely unknown to the sender.\\
To quantify the mixedness of the state (\ref{rho1inex}), we
calculate the linear entropy $S_{L}$ \cite{munro}
\begin{eqnarray}
&&S_{L}= \frac{4}{3}(1-Tr([\rho_{1}^{in}])^{2})=
\frac{8}{3}(\frac{1}{4}-|y|^{2}) \label{linentr}
\end{eqnarray}
We note here that for lower values (in the neighborhood of zero)
of $|y|$ the input qubit (\ref{rho1inex})
is maximally mixed. As the values of $|y|$ proceed towards $\frac{1}{2}$ the mixedness of the input qubit decreases.\\
For the above class of input states we find that the Hilbert
Schmidt norms $D_{1}$ and $D_{2}$ are equal and given by
\begin{eqnarray}
&&D_{12}= 2(1-p_{1})^{2}(\frac{1}{4}+|y|^{2}),~~~0\leq |y| \leq
\frac{1}{2} \label{dist2}
\end{eqnarray}\\
\textbf{CASE-III}: When $rank([\rho_{23}]^{MEMS})=3$ i.e., when
$p_{4}=0$. Without any loss of generality, we assume
$p_{1}=p_{2}$. Therefore the teleportation channel
$[\rho_{23}]^{MEMS}$ reduces to
\begin{eqnarray}
[\rho_{23}]^{MEMS}_{r3}= \left(\begin{matrix}{\frac{1-p_{1}}{2} &
0 & 0 & \frac{1-3p_{1}}{2} \cr 0 & p_{1} & 0 & 0 \cr 0 & 0 & 0 & 0
\cr \frac{1-3p_{1}}{2} & 0 & 0 &
\frac{1-p_{1}}{2}}\end{matrix}\right) \label{rho23r3}
\end{eqnarray}
The maximum concurrence of the state $[\rho_{23}]^{MEMS}_{r3}$ is $C([\rho_{23}]^{MEMS}_{r3})=3p_{1}-1$.\\
If Alice uses $[\rho_{23}]^{MEMS}_{r3}$ as a teleportation channel
to teleport an unknown qubit described by the density matrix
(\ref{rho1ex}) then the state after Alice's measurement and Bob's
unitary transformation (like in case-II) is given by
\begin{eqnarray}
&&[\rho^{B_{5}}_{3}]=
\frac{1}{N}\left(\begin{matrix}{\frac{x(1-p_{1})}{2} &
\frac{y(3p_{1}-1)}{2} \cr \frac{y^{*}(3p_{1}-1)}{2} &
xp_{1}+\frac{(1-x)(1-p_{1})}{2}}\end{matrix}\right) ~\textrm{and}~
{}\nonumber\\&& [\rho^{B_{6}}_{3}]=
\frac{1}{N_{1}}\left(\begin{matrix}{\frac{x(1-p_{1})}{2}+(1-x)p_{1}
&  \frac{y(3p_{1}-1)}{2} \cr \frac{y^{*}(3p_{1}-1)}{2} &
\frac{(1-x)(1-p_{1})}{2}}\end{matrix}\right) \label{rhoB56}
\end{eqnarray}
where $N=x(\frac{1+p_{1}}{2})+(1-x)(\frac{1-p_{1}}{2})$ and $N_{1}=x(\frac{1-p_{1}}{2})+(1-x)(\frac{1+p_{1}}{2})$.\\
The Hilbert Schmidt norm of the difference between two density
matrices $[\rho_{1}]$ at Alice's side and $[\rho^{B_{5}}_{3}]~~
\textrm{or}~~ [\rho^{B_{6}}_{3}]$ (corresponding to different
measurements) at Bob's side is given by
\begin{eqnarray}
&&D_{3}= Tr([\rho_{1}]-[\rho_{3}^{B_{5}}])^{2}=
\frac{2}{N^{2}}(x^{4}p_{1}^{2}+|y|^{2}(1-2p_{1}+p_{1}x)^{2})~~~~\textrm{and}~~~~
{}\nonumber\\&& D_{4}= Tr([\rho_{1}]-[\rho_{3}^{B_{6}}])^{2}=
\frac{2}{N^{2}_{1}}((1-x)^{4}p_{1}^{2}+|y|^{2}(1-p_{1}-xp_{1})^{2})
\label{dist4}
\end{eqnarray}
Interestingly we find that for the same class of input state
(\ref{rho1inex}), the Hilbert Schmidt norms $D_{3}$ and $D_{4}$
are equal and given by
\begin{eqnarray}
&&D_{34}=
4(\frac{p_{1}^{2}}{8}+2|y|^{2}(1-\frac{3p_{1}}{2})^{2}),~~~0\leq
|y| \leq \frac{1}{2} \label{dist5}
\end{eqnarray}\\
\textbf{CASE-IV}: When $rank([\rho_{23}]^{MEMS})=4$ i.e., when
$p_{4}\neq 0$. Without any loss of generality, we assume
$p_{2}=p_{3}=p_{4}$. Therefore the teleportation channel
$[\rho_{23}]^{MEMS}$ reduces to
\begin{eqnarray}
[\rho_{23}]^{MEMS}_{r4}= \left(\begin{matrix}{\frac{1+2p_{1}}{6} &
0 & 0 & \frac{1-4p_{1}}{6} \cr 0 & \frac{1-p_{1}}{3} & 0 & 0 \cr 0
& 0 & \frac{1-p_{1}}{3} & 0 \cr \frac{1-4p_{1}}{6} & 0 & 0 &
\frac{1+2p_{1}}{6}}\end{matrix}\right) \label{rho23r4}
\end{eqnarray}
The maximum concurrence of the state $[\rho_{23}]^{MEMS}_{r4}$ is $C([\rho_{23}]^{MEMS}_{r4})=2p_{1}-1$.\\
If Alice uses $[\rho_{23}]^{MEMS}_{r4}$ as a teleportation channel
to teleport an unknown qubit described by the density matrix
(\ref{rho1ex}) then interestingly we find that corresponding to
each of Alice's measurement outcome
$(|\Phi^{+}>,|\Phi^{-}>,|\Psi^{+}>, |\Psi^{-}>)$ and Bob's unitary
operation $(\sigma_{z}, I, \sigma_{y}, \sigma_{x})$, the state in
Bob's side reduces to
\begin{eqnarray}
&&[\rho^{B_{7}}_{3}]=
\left(\begin{matrix}{\frac{x(2p_{1}+1)}{3}+\frac{2(1-x)(1-p_{1})}{3}
& \frac{y(1-4p_{1})}{3} \cr \frac{y^{*}(4p_{1}-1)}{3} &
\frac{2x(1-p_{1})}{3}+\frac{(1-x)(1+2p_{1})}{3}}\end{matrix}\right)
\label{rhoB7}
\end{eqnarray}
In this case the Hilbert Schmidt norm is given by
\begin{eqnarray}
&&D_{5}=
Tr([\rho_{1}]-[\rho_{3}^{B_{7}}])^{2}=\frac{8(1-p_{1})^{2}((2x-1)^{2}+4|y|^{2})}{9}
\label{dist7}
\end{eqnarray}
For the class of input states given in (\ref{rho1inex}), the eqn.
(\ref{dist7}) reduces to
\begin{eqnarray}
&&D_{56}= \frac{32(1-p_{1})^{2}|y|^{2}}{9},~~~0\leq |y| \leq
\frac{1}{2} \label{dist8}
\end{eqnarray}
It is to be noted that the family of Werner states of the form
\begin{eqnarray}
[\rho_{23}]^{W}= \left(\begin{matrix}{\frac{1+r}{4} & 0 & 0 &
\frac{-r}{2} \cr 0 & \frac{1-r}{4} & 0 & 0 \cr 0 & 0 &
\frac{1-r}{4} & 0 \cr \frac{-r}{2} & 0 & 0 &
\frac{1+r}{4}}\end{matrix}\right) \label{rho23w}
\end{eqnarray}
falls under this category because the rank of $[\rho_{23}]^{W}$ is
also 4.\\
Comparing the density matrix $[\rho_{23}]^{MEMS}_{r4}$ with
$[\rho_{23}]^{W}$, we find the relationship between the parameters
$r$ and $p_{1}$ as
\begin{eqnarray}
p_{1}=\frac{1+3r}{4}\label{relationship}
\end{eqnarray}
In this case, the distortion $D_{56}$ given in (\ref{dist8})
reduces to
\begin{eqnarray}
D_{56}^{W}= 2(1-r)^{2}|y|^{2}\label{wernerdist.}
\end{eqnarray}
Since the distortion $D_{56}^{W}$ depends on the input state
parameter so averaging over all input states, we get
\begin{eqnarray}
\overline{D}_{56}^{W}= \frac{(1-r)^2}{12}\label{avgwernerdist.}
\end{eqnarray}
Now we can speculate few cases:\\
(i) If $r=1$, then the teleportation channel would be maximally
entangled pure state and as expected, the distortion of the
output state from the input state is zero.\\
(ii) If $r=0$, then the teleportation channel would be maximally
mixed state and hence the average distortion is maximum in this
case and it is found out to be $\frac{1}{12}$.\\
(iii) If $0<r<1$ then the quantum teleportation channel would be
any channel lying between maximally entangled pure state and
maximally mixed state. In this case, the average distortion is
lying between $0$ and $\frac{1}{12}$.
\subsection{Graphical study of the teleportation scheme }
In this section we analyze (i) graph for Hilbert Schmidt distance
$D$ against input state parameter $|y|$ for different values of
teleportation channel parameter $p_{1}$ and (ii) graph for linear
entropy $S_{L}$ versus teleportation channel parameter $p_{1}$.\\\\
We make some interesting observations from the above graphs:\\
$\textbf{G1}.$ The Hilbert Schmidt distances $D_{12}$, $D_{34}$
and $D_{56}$
are increasing function of $|y|^{2}$ for any given $p_{1}$.\\
$\textbf{G2-I.}$ When $0\leq p_{1}\leq\frac{1}{2}$, the ordering
of the Hilbert Schmidt distances $D_{12}$ and $D_{34}$ is given by
(i) $D_{34}\leq D_{12}$, if $0\leq
|y|^{2}\leq\frac{1-2p_{1}}{4(8p_{1}^{2}-4p_{1}+3)}$ and (ii)
$D_{12}<D_{34}$, if $\frac{1-2p_{1}}{4(8p_{1}^{2}-4p_{1}+3)}<
|y|^{2}\leq \frac{1}{4}$. The distances $D_{12}$ and $D_{34}$ are
equal only when $|y|^{2}=\frac{1-2p_{1}}{4(8p_{1}^{2}-4p_{1}+3)}$.
In the graph the intersection point of the curves for $D_{12}$ and
$D_{34}$ is shown at P.\\
$\textbf{G2-II}.$ When $\frac{1}{2}< p_{1}< 1$, the ordering of
$D_{12}$, $D_{34}$ and $D_{56}$ is given by
$D_{56}<D_{12}<D_{34}$. This observation provide us a clue to
arrive at a conclusion that the two qubit state described by the
density matrices of rank-4 will be the better option to opt as a
teleportation channel when the channel parameter $p_{1}$ is
greater than $\frac{1}{2}$.\\
$\textbf{G3-I.}$ As we increase the value of $p_{1}$, the minimum
value of $D_{12}$ is attained at $|y|=0$ and this minimum value is
shifted downward along vertical axis. This shows that the
teleportation of the single qubit mixed state will be better for
higher values of $p_{1}$ (i.e., for those values of $p_{1}$ which
lies in the neighborhood of unity) via
the two qubit state described by rank-2 density matrices as a teleportation channel.\\
$\textbf{G3-II.}$ As we increase the value of $p_{1}$ starting
from $\frac{1}{3}$, the minimum value of $D_{34}$ is also attained
at $|y|=0$ but this minimum value is shifted upward along vertical
axis. This means that the teleportation of the single qubit mixed
state will be worse for higher values of $p_{1}$
when we use the two qubit state described by rank-3 density matrices as a teleportation channel.\\
$\textbf{G3-III.}$ The minimum value of $D_{56}$ remain fixed at $|y|=0$ for $p_{1}> \frac{1}{2}$.\\
$\textbf{G4.}$ Let $S_{L}^{r_{2}}$, $S_{L}^{r_{3}}$ and
$S_{L}^{r_{4}}$ denote the linear entropies of  two qubit
teleportation channel described by the density matrix of rank-2,
rank-3 and rank-4 respectively.\\
(i) When $p_{1}<\frac{1}{2}$, the ordering of the linear entropies
is given by $S_{L}^{r_{2}}<S_{L}^{r_{3}}<S_{L}^{r_{4}}$.\\
(ii)When $p_{1}>\frac{1}{2}$, the ordering of the linear entropies
is given by $S_{L}^{r_{3}}<S_{L}^{r_{2}} < S_{L}^{r_{4}}$.\\
We note that the curve for the linear entropies $S_{L}^{r_{2}}$
and $S_{L}^{r_{3}}$ intersect at R $(0.5,0.66)$.\\
It is therefore clear that the mixedness of the two qubit
teleportation channel described by the density matrix of rank-4 is
greater than the mixedness of the density matrix of rank-2 and
rank-3 respectively for any value of the parameter $p_{1}$.
\section{Conclusion}
We have studied partial trace and partial transposition in terms
of block matrices and thereafter derived an entanglement condition
for bipartite $2\otimes2$ system. Also we have presented a
mathematical formulation of the original teleportation scheme. In
the teleportation scheme, we consider a particular set of single
qubit mixed state which is to be teleported. Then we find that the
success of the teleportation depends on the channel parameter
$p_{1}$ i.e. when $p_{1}<\frac{1}{2}$, the two qubit teleportation
channel described by the density matrices of rank-2 or rank-3 will
be useful to teleport a single qubit mixed state described by the
parameter $\frac{1-2p_{1}}{4(8p_{1}^{2}-4p_{1}+3)}< |y|^{2}\leq
\frac{1}{4}$ or $0\leq
|y|^{2}\leq\frac{1-2p_{1}}{4(8p_{1}^{2}-4p_{1}+3)}$. When
$p_{1}>\frac{1}{2}$, the maximally entangled mixed two qubit
teleportation channel described by the density matrix of rank-4
should be used to teleport a single qubit mixed state.

\newpage
\begin{figure}
\vbox{
\centering{
\hskip -5cm
\vskip -3cm
\psfig{figure=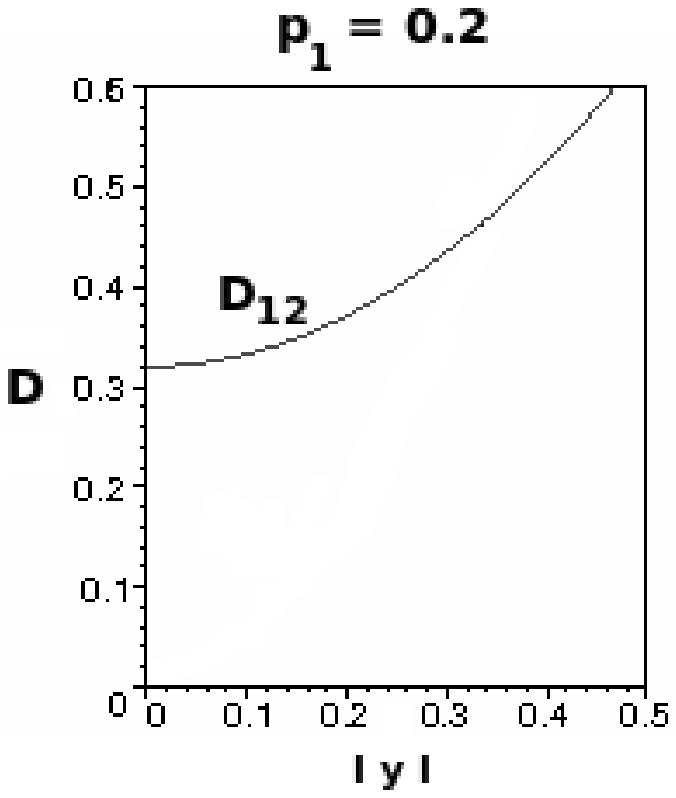,width=10truecm}}}
\caption{The Hilbert Schmidt Norm $D_{12}$ between the input state and the teleported state
 via quantum channel described by the density matrix of rank-2 is plotted versus the input state
 parameter $|y|$ when $p_{1}=0.2$ }
\end{figure}

\begin{figure}
\vbox{
\centering{
\hskip-0.5cm
\psfig{figure=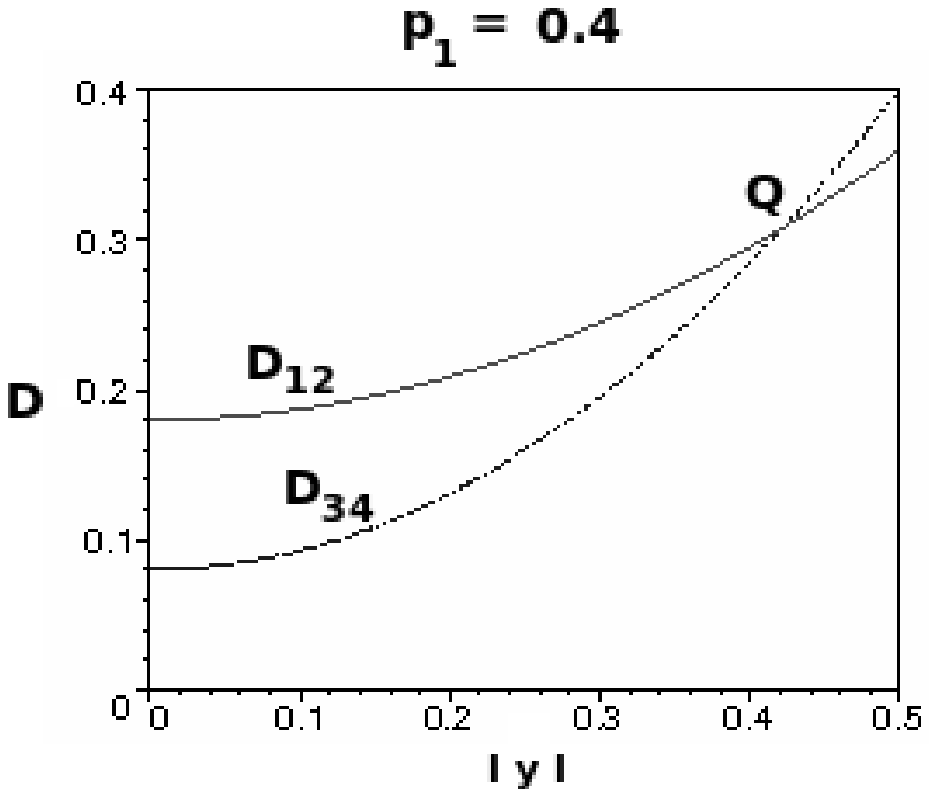,width=10truecm}}}
\caption{The Hilbert Schmidt Norms $D_{12},D_{34}$  between the input state and the teleported state
 via quantum channel described by the density matrix of rank-2 and rank-3 is plotted versus the input state
 parameter $|y|$ when $p_{1}=0.4$}
\end{figure}

\begin{figure}
\vbox{
\centering{
\hskip-0.5cm
\psfig{figure=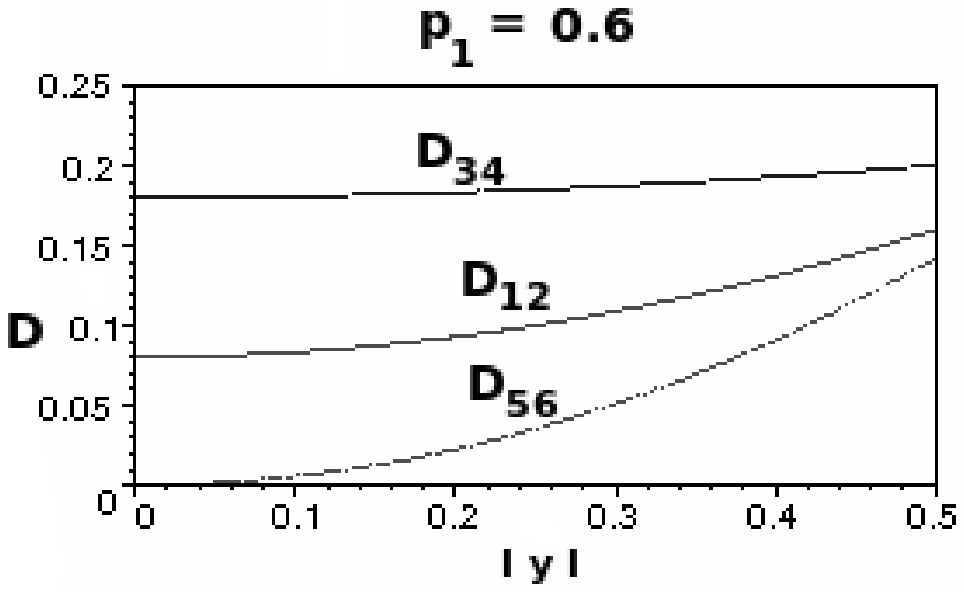,width=10truecm}}}
\caption{The Hilbert Schmidt Norms $D_{12},D_{34},D_{56}$ between the input state and the teleported state
 via quantum channel described by the density matrix of rank-2, rank-3 and rank-4 is plotted versus the input state
 parameter $|y|$ when $p_{1}=0.6$}
\end{figure}

\begin{figure}
\vbox{
\centering{
\psfig{figure=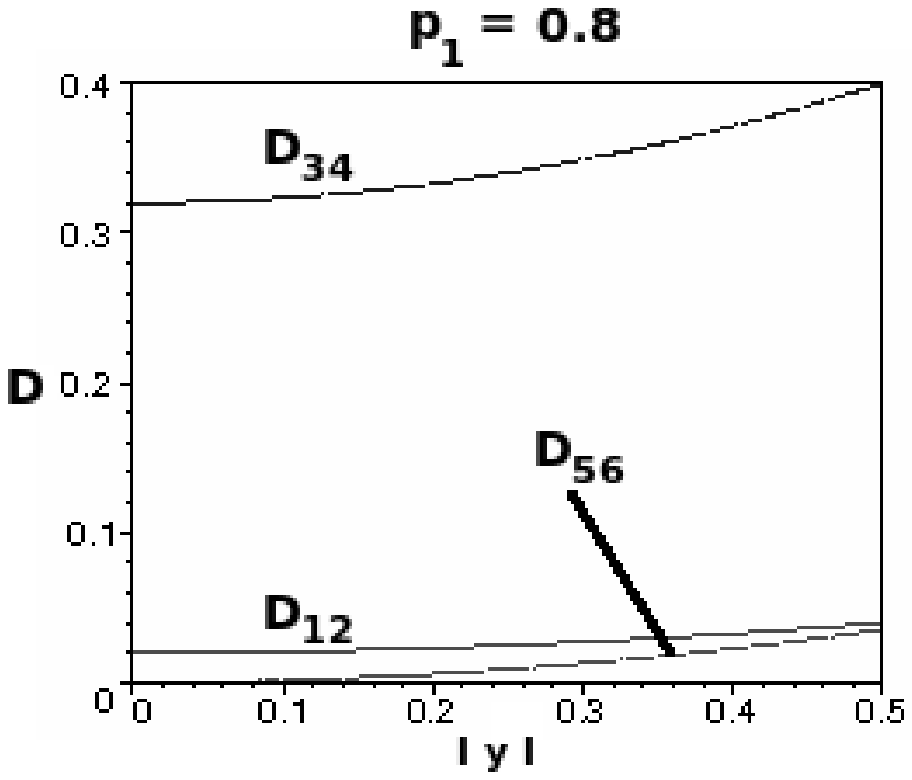,width=10truecm}}}
\caption{The Hilbert Schmidt Norms $D_{12},D_{34},D_{56}$ between the input state and the teleported state
 via quantum channel described by the density matrix of rank-2, rank-3 and rank-4 is plotted versus the input state
 parameter $|y|$ when $p_{1}=0.8$}
\end{figure}

\begin{figure}
\vbox{
\centering{
\hskip-0.5cm
\psfig{figure=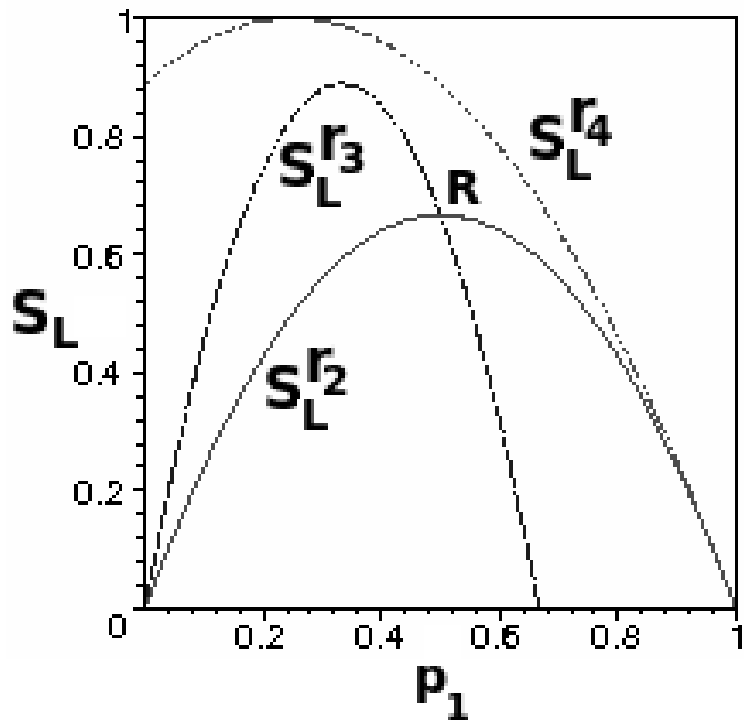,width=10truecm}}}
\caption{The linear entropies $S_{L}^{r_{2}},S_{L}^{r_{3}},S_{L}^{r_{4}}$
 for two qubit teleportation channel described by the density matrix of rank-2, rank-3 and rank-4  are plotted
 versus the channel parameter $p_{1}$ }
\end{figure}

\end{document}